\documentclass[]{acta}
\usepackage{lscape}
\usepackage{amssymb}
\usepackage{amsmath}
\usepackage[T1]{fontenc}
\usepackage{lmodern}
\usepackage[english]{babel}
\usepackage[autostyle]{csquotes}
\usepackage{enumerate}
\usepackage{xspace}
\usepackage{graphicx}
\usepackage{amssymb}
\usepackage{amsmath}
\usepackage{lscape}

\def\url#1{\expandafter\string\csname #1\endcsname}

\newcommand{\gaia}{GDR2\xspace}
\newcommand{\pan}{PDR1\xspace}
\newcommand*\arcsec{\ensuremath{^{\prime\prime}}}

\newcommand{\mud}{\mu_{\delta}}

\newcommand*\aap{A\&A}

\newcommand*\actaa{Acta Astron.}
\newcommand*\aj{AJ}
\newcommand*\apj{ApJ}
\newcommand*\apjl{ApJ}

\newcommand*\apjs{ApJS}

\newcommand*\mnras{MNRAS}
\newcommand*\nat{Nature}
\newcommand*\pasp{PASP}

\SetPages{0}{0}

\SetVol{58}{2008}

\begin{document}

\begin{Titlepage}
\Title{Predicted microlensing events by nearby very-low-mass objects:\\ Pan-STARRS DR1 vs. \textit{Gaia} DR2}
\Author{M.B. Nielsen$^1$ and D.M. Bramich$^2$}
{$^1$Center for Space Science, NYUAD Institute, New York University Abu Dhabi, PO Box 129188, Abu Dhabi, UAE\\
e-mail:mbn4@nyu.edu\\
$^2$New York University Abu Dhabi, PO Box 129188, Saadiyat Island, Abu Dhabi, UAE\\
e-mail:dan.bramich@hotmail.co.uk}

\Received{Month Day, Year}
\end{Titlepage}

\Abstract{Microlensing events can be used to directly measure the masses of single field stars to a precision of $\sim$1-10\%. The majority of direct mass measurements for stellar and sub-stellar objects typically only come from observations of binary systems. Hence microlensing provides an important channel for direct mass measurements of single stars. The \textit{Gaia} satellite has observed $\sim$1.7~billion objects, and analysis of the second data release has recently yielded numerous event predictions for the next few decades. However, the \textit{Gaia} catalog is incomplete for nearby very-low-mass objects such as brown dwarfs for which mass measurements are most crucial. We employ a catalog of very-low-mass objects from Pan-STARRS data release 1 (PDR1) as potential lens stars, and we use the objects from \textit{Gaia} data release 2 (\gaia) as potential source stars. We then search for future microlensing events up to the year 2070. The Pan-STARRS1 objects are first cross-matched with \gaia to remove any that are present in both catalogs. This leaves a sample of $1,718$ possible lenses. We fit MIST isochrones to the Pan-STARRS1, AllWISE and 2MASS photometry to estimate their masses. We then compute their paths on the sky, along with the paths of the \gaia source objects, until the year 2070, and search for potential microlensing events. Source-lens pairs that will produce a microlensing signal with an astrometric amplitude of greater than 0.131~mas, or a photometric amplitude of greater than 0.4~mmag, are retained.}
{gravitational lensing: micro -- methods: data analysis -- catalogs -- astrometry -- stars: fundamental parameters}

\section{Introduction}
As a compact massive object moves close to the line-of-sight with a background source star, due to either parallax and/or proper motion, the observer will start to view two distorted and magnified images of the source (Einstein 1915; Einstein 1936; Liebes 1964; Refsdal 1964). The two images are generally unresolvable with current observing facilities. The observed global source magnification is termed photometric microlensing (Beaulieu et al. 2006), and the observed deflection of the apparent location of the source centroid is termed astrometric microlensing (Sahu et al. 2017). 

When the distances to the source and the lens are already known, the mass of the lensing object may be determined by observing the astrometric signal alone. However, observation of both the photometric and astrometric signals when present provides even stronger lens mass constraints. The mass estimates derived from observations of microlensing events are independent of any assumptions about the internal stellar physics, and they provide an opportunity to measure the masses of single stars for which so few measurements exist. While photometric microlensing events are routinely observed by, e.g., the Optical Gravitational Lensing Experiment (OGLE, Udalski 2003; Udalski, Szyma\'nski, and Szyma\'nski 2015; Wyrzykowski et al. 2015) or the Microlensing Observations in Astrophysics survey (MOA, Sumi et al. 2013), the distances to the source and lens are, however, often difficult to measure using ground-based observatories.

\textit{Gaia} data release 2 (\gaia; Prusti et al. 2016; Gaia Collaboration, Brown et al. 2018) has alleviated this problem by providing exceptionally precise measurements of parallax and proper motions for more than $\sim$1.3 billion stars over the whole sky. This has been exploited by Bramich (2018) and Bramich and Nielsen (2018) to predict 76 and 2,509 microlensing events, respectively, during the next decade and until the end of the 21st century. However, \gaia is by no means complete in terms of optically faint, very red, nearby targets with high proper motions. Gaia Collaboration, Brown et al. (2018) estimate that $\sim17\%$ of stars with proper motion $>0.6\arcsec \mathrm{yr}^{-1}$ are missing in \gaia. Nearby very-low-mass (VLM) objects such as late M-, L- and T-type dwarfs fall into this category.  

VLM objects are particularly interesting for exoplanet searches (e.g., the MEarth project; Irwin et al. 2015) as they are amenable to follow-up investigations by moderate-sized telescopes. In the event that an exoplanet host produces one or more microlensing events, its mass may be measured, which in turn provides estimates of the masses of any orbiting companions. Moreover, estimating the masses of VLM objects is especially important as these ultra-cool dwarfs straddle the divide between stars and giant planets, and thus provide constraints for stellar evolutionary models and classification efforts for the lower part of the main sequence (e.g., Burrows, Sudarsky, and Hubeny 2006; Burgasser, Burrows, and Kirkpatrick 2006). 

Here we use a sample of VLM objects compiled by Best, Magnier, et al. (2018), which were observed during the Pan-STARRS1 $3\pi$ survey  (Kaiser et al. 2010; Chambers et al. 2016). This sample of VLM objects is used as a catalog of lenses while the \gaia is used as a catalog of sources, in order to predict potential future microlensing events over the whole sky up until the year 2070. 

\section{Methods}
Predicting microlensing events essentially requires one to compute the future paths of the stars in the VLM sample, and compare these to the future paths of the stars in \gaia. This is obviously a computationally intensive task considering the enormity of the \gaia, even when compared to a relatively modest sample of VLM objects in the \pan. We therefore take a multi-step approach to minimize the number of source-lens pairs that we need to consider in detail. The methods used to identify source-lens pairs from the VLM sample and from \gaia that could potentially lead to microlensing events follow those described in Bramich (2018). Hence we only cover the methods in detail where necessary, and we refer the reader to Bramich (2018) if further information is required.

\subsection{VLM objects in Pan-STARRS1 DR1 as lenses}
\begin{figure*}
\centering
\includegraphics[width = \columnwidth]{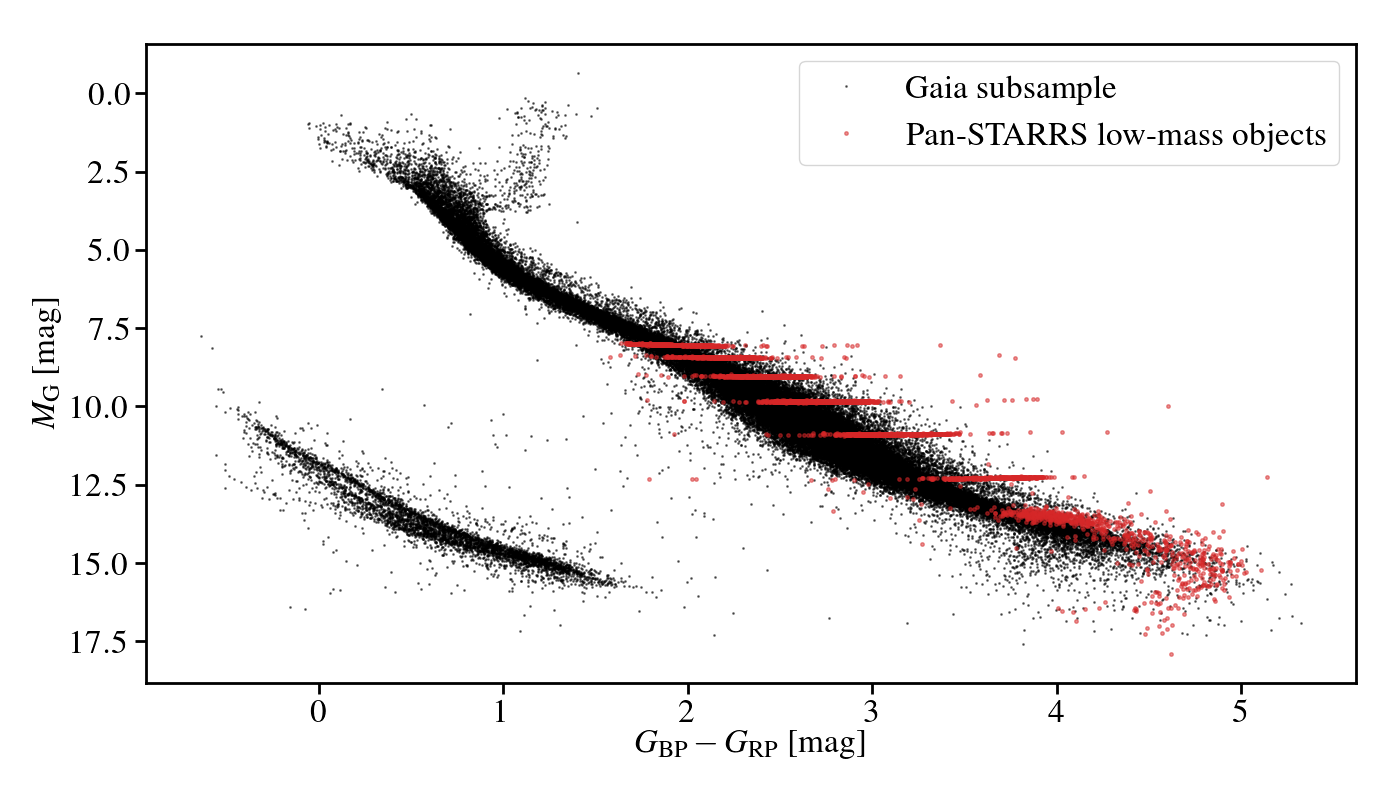}
\caption{Color-magnitude diagram in the \textit{Gaia} $G_{\mbox{\scriptsize BP}}$, $G_{\mbox{\scriptsize RP}}$ and $G$ bands. VLM objects from \pan are shown in red. Conversions to \textit{Gaia} passbands are taken from Jordi et al. (2010); this limits us to only showing objects that have been observed in $g_{\mbox{\scriptsize PS1}}$, $r_{\mbox{\scriptsize PS1}}$, and $i_{\mbox{\scriptsize PS1}}$ ($6,157$ targets). The discretization of the $M_{G}$ values most clearly seen above $M_{G}\approx 12.5$~mag is a consequence of the photometric distances (used for computing absolute magnitudes) being derived from stellar models of discrete spectral types (Best, Magnier, et al. 2018). Note that none of the VLM lens stars for which we make predictions in this work have $g_{\mbox{\scriptsize PS1}}$-band photometry and so they are not displayed here. A sub-sample of randomly selected \gaia stars are shown in black for reference, following the selection criteria in Section 2 of Gaia Collaboration, Babusiaux (2018). }
\label{fig:HR}%
\end{figure*}

The VLM sample by Best, Magnier, et al. (2018) comprises almost all of the currently known L0-T2 dwarf stars and also includes some additional late M-dwarf stars as well; although the M-dwarf sample is incomplete.

The catalog of VLM objects contains astrometry and photometry for $9,888$ stars. The photometry is available in the following passbands: Pan-STARRS1\footnote{Processing Version 3.3} $g$, $r$, $i$, $z$ and $y$; 2MASS $J$, $H$, and $Ks$; AllWISE $W1$, $W2$, $W3$, and $W4$; and the \textit{Gaia} $G$-band. Note that not all stars have photometric measurements in all passbands. We use the broad-band photometry along with isochrone fitting to provide an initial estimate of the stellar mass, which is used to identify source-lens pairs that have the potential to  produce microlensing events (see below). Figure \ref{fig:HR} shows a color-magnitude diagram of a selection of the VLM objects along with a subsample from \gaia. A small fraction of the VLM objects from \pan do not have photometry in the optical passbands, and their NIR photometry is not easily converted into the \textit{Gaia} blue ($G_{\mbox{\scriptsize BP}}$) and red ($G_{\mbox{\scriptsize RP}}$) passbands. We have therefore omitted these objects from Fig. 1. 

Right-ascension, $\alpha$, and declination, $\delta$, are available for the entire sample of $9,888$ stars (with a different reference epoch for each object). However, only $9,763$ stars have proper motions $\mu_{\alpha*}$ and $\mu_{\delta}$, and only $9,654$ of these have either a parallax $\varpi$ measurement (taken from the literature) or a photometric distance. Best, Magnier et al. (2018) compute the photometric distances by comparing the $W2$ photometry with absolute $W2$ magnitudes from absolute magnitude - spectral type polynomials by Dupuy and Liu (2012). We invert the photometric distances to parallaxes along with their uncertainties, and incorporated them in the prediction computations in an identical manner to the directly-measured parallaxes. 

We keep the $9,654$ stars that have all five astrometric parameters $\alpha$, $\delta$, $\mu_{\alpha*}$, $\mu_{\delta}$ and $\varpi$ as our set of potential lens stars.

\subsection{\gaia stars as background sources}

For the background source stars, we use exactly the same sample that was selected from \gaia in Bramich (2018) (their Section~4.2.3). This sample consists of $1,366,072,323$ stars. Note that \gaia quantities for the source stars are corrected; namely, $0.029$~mas is added to the parallaxes, and the uncertainties on the astrometric parameters are inflated by $25\%$ (Bramich 2018).

\subsection{Initial source-lens pair selection}
We perform the initial source-lens pair selection by computing a conservative upper limit $\theta_{\mbox{\scriptsize E,max}}$ on the value of the Einstein radius $\theta_{\mbox{\scriptsize E}}$ for any particular source-lens pair by assuming a maximum lens mass of $M_{\mbox{\scriptsize max}}=2M_{\odot}$, a maximum lens parallax of $\varpi_{\mbox{\scriptsize L,max}}=\varpi_{\mbox{\scriptsize L}}+3\sigma[\varpi_{\mbox{\scriptsize L}}]$ (where $\sigma[\varpi_{\mbox{\scriptsize L}}]$ is the uncertainty on the lens parallax $\varpi_{\mbox{\scriptsize L}}$), and a source parallax of zero. This translates into a maximum source-lens angular separation $\theta_{\mbox{\scriptsize det}}$ within which a microlensing signal can be detected by considering the astrometric deflection and the best astrometric precision achievable by any current observing facility in a single observation (excluding radio interferometry). The astrometric deflection is used as this has a larger range of influence than the photometric magnification (Dominik and Sahu 2000; Belokurov and Evans 2002). One obtains $\theta_{\mbox{\scriptsize det}} =  \theta_{\mbox{\scriptsize E,max}}^{\,2} / 0.030$~mas, where 0.030~mas is the bright-limit along-scan astrometric precision for \textit{Gaia} (Rybicki et al. 2018; Bramich 2018).

We calculate $\theta_{\mbox{\scriptsize det}}$ for each potential lens star. To account (very) conservatively for source and lens motions, and for errors, $\sigma$, in the astrometric parameters, we compute the following quantity for each source-lens (S, L) pair:
\begin{equation}
\begin{aligned}
\theta^{\prime}_{\mbox{\scriptsize det}} = & \,\, \theta_{\mbox{\scriptsize det}}
                                             + 3\sigma[\alpha*_{\mbox{\scriptsize ref,S}}] 
                                             + 3\sigma[\delta_{\mbox{\scriptsize ref,S}}] \\
                                           & + T_{\mbox{\scriptsize S}} \, \left( \,|\,\mu_{\alpha*,\mbox{\scriptsize S}}| + 3\sigma[\mu_{\alpha*,\mbox{\scriptsize S}}]
                                                                          + |\,\mu_{\delta,\mbox{\scriptsize S}}| + 3\sigma[\mu_{\delta,\mbox{\scriptsize S}}] \,\right) \\
                                           & + T_{\mbox{\scriptsize L}} \, \left( \, |\,\mu_{\alpha*,\mbox{\scriptsize L}}| + 3\sigma[\mu_{\alpha*,\mbox{\scriptsize L}}]
                                                                          + |\,\mu_{\delta,\mbox{\scriptsize L}}| + 3\sigma[\mu_{\delta,\mbox{\scriptsize L}}] \,\right)   \\
                                           & + \varpi_{\mbox{\scriptsize S}} + 3\sigma[\varpi_{\mbox{\scriptsize S}}]
                                             + \varpi_{\mbox{\scriptsize L}} + 3\sigma[\varpi_{\mbox{\scriptsize L}}],                                                      \\
\label{eqn:newlim}
\end{aligned}
\end{equation}
where $\alpha_\mathrm{ref}$ and $\delta_\mathrm{ref}$ are the celestial coordinates at the reference epoch. The uncertainty on $\alpha$, as well as the proper motion $\mu_\alpha$ and its uncertainty are corrected by a factor $\cos{\delta_{\mathrm{ref}}}$ for projection effects, denoted by *.

Note that uncertainties on $(\alpha_{\mbox{\scriptsize ref,L}},\delta_{\mbox{\scriptsize ref,L}})$ are not available in Best, Magnier, et al. (2018) and the relevant terms are therefore not included in the above expression.  $T_{\mbox{\scriptsize S}}=54.5$~years is the length of time from the \gaia reference epoch J2015.5 until J2070.0, and $T_{\mbox{\scriptsize L}}$ is the length of time from the PDR1 reference epoch for the star in question until J2070.0. An error on $(\alpha_{\mbox{\scriptsize ref,L}},\delta_{\mbox{\scriptsize ref,L}})$ introduces exactly the same error on the closest approach of a source-lens pair. Chambers et al. (2016) estimate an error of $\sim2$~mas on the \pan $\alpha_{\mbox{\scriptsize ref,L}}$ and $\delta_{\mbox{\scriptsize ref,L}}$. Errors of this magnitude may change the shape of the photometric and astrometric signals, but we do not expect a significant impact on their amplitudes.

We reject all source-lens pairs for which the angular distance between their coordinates at their respective reference epochs exceeds $\theta^{\prime}_{\mbox{\scriptsize det}}$. This leaves $27,966$ source-lens pairs, with $6,157$ unique lenses, for further consideration.

\subsection{Removing source-lens pairs for stars in both PDR1 and \gaia}
\label{sec:removing}

\begin{figure}
\centering
\includegraphics[width=\linewidth]{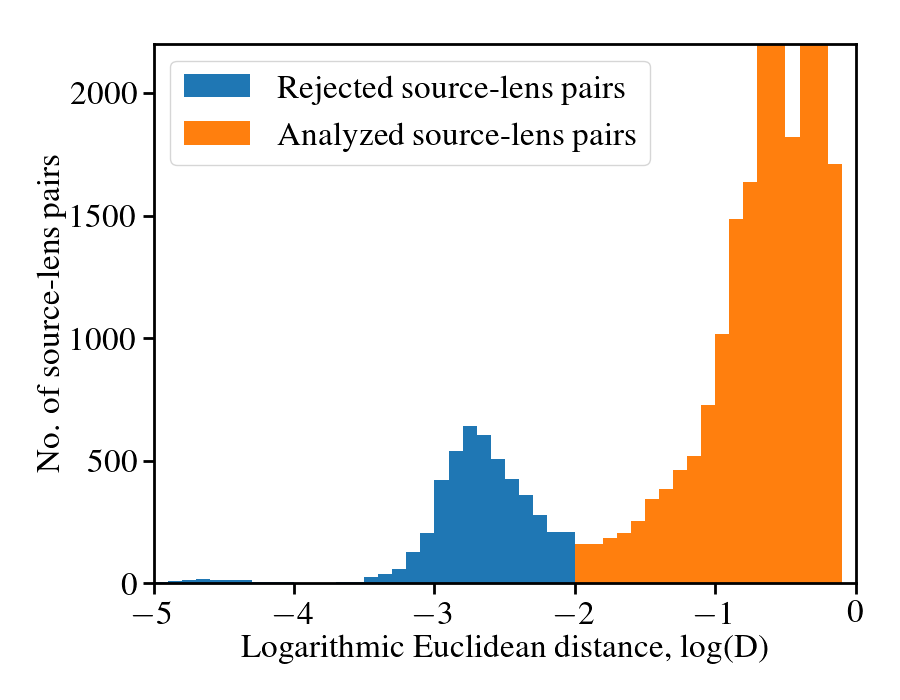}
\caption{Five dimensional distances between \textit{Gaia} source and Pan-STARRS1 lens pairs. Pairs with a distance $\log(D) < -2$ are assumed to be either (i) the same object represented in both catalogs, or (ii) members of a binary or co-moving system.}.
\label{fig:eucl}
\end{figure}

It is expected that some objects from \pan are also present in \gaia. This raises the issue that some of the lenses in our lens sample may be inadvertently paired with a source from \gaia that is in fact the same star. Bramich (2018) and Bramich and Nielsen (2018) have already performed an analysis of \gaia as a catalog of both source and lens objects. Hence any lens star from our sample that is also in \gaia has already been fully considered for producing microlensing events.

To avoid this issue, and further reduce the number of possible source-lens pairs that we need to consider, we initially compute the distance $D$ between the source and the lens in each source-lens pair using the astrometric parameter differences $\Delta\alpha*_{\mathrm{ref}}$, $\Delta\delta_{\mathrm{ref}}$, $\Delta\mu_{\alpha*}$, $\Delta\mud$, $\Delta\varpi$. Here $\alpha$ and $\delta$ for the lens stars have been adjusted using their proper motions and their PDR1 reference epoch to match the \textit{Gaia} reference epoch of J2015.5. Source-lens pairs with $\log(D) < -2$ are then rejected. This limit is imposed based on the clear separation of the computed distances into two well-defined groups as shown in Fig. \ref{fig:eucl}. In terms of $\alpha$ and $\delta$, this filter corresponds to removing almost all source-lens pairs that are within $1\arcsec$ of each other at J2015.5, i.e., that are very likely the same objects. Any remaining pairs with angular separations of less than $1\arcsec$ have differing proper motions and/or parallax. Stars in comoving or binary pairs are expected to have similar proper motion, parallax, and location, and so should be removed by this filtering. From this filtering step, we are left with 23,253 source-lens pairs with 1,718 unique lenses.

\subsection{Lens mass estimation}
\label{sec:mass}
To compute a mass estimate for each lens we use the Isochrones (Morton 2015) package\footnote{https://github.com/timothydmorton/isochrones} for Python. This takes a number of photometric observations in different passbands and finds the best fit stellar model in the MESA Isochrones \& Stellar Tracks grid (MIST; Dotter 2016; Choi et al. 2016). Here we use all of the available passbands (see above) that Isochrones allows with the MIST grid. Each fit is done using an MCMC sampler\footnote{http://dfm.io/emcee/current/} (Foreman-Mackey et al. 2013), and produces a posterior distribution of the mass of each lens star. We adopt the median of the posterior distribution as the estimate of the lens mass in each case. 

We note that this model grid is not optimal for brown dwarfs, and that it has a lower mass limit of $M = 0.1M_{\odot}$, while some of the lenses in our sample will likely have masses around $M\sim0.08M_{\odot}$. In our fits this produces a pile-up of best-fit masses immediately above $0.1M_{\odot}$. However, even a $\sim25\%$ error in the mass estimate is still too small to substantially affect the microlensing predictions presented in this paper. Errors in the mass estimates do not affect our predictions of when an event will peak. As long as a predicted event still produces a detectable signal with approximately the same signal amplitude, then our mass estimates are sufficient.  

\subsection{Identifying future microlensing events}
\label{sec:monte}

For each of the $23,253$ source-lens pairs from Section~\ref{sec:removing}, we perform $1,000$ Monte Carlo simulations of the source and lens paths on the sky. Each simulation is generated
as follows:
\begin{enumerate}[(i)]
\item{We draw a set of proper motion and parallax parameters for the lens star from Gaussian distributions defined by the lens proper motion and parallax values and their uncertainties. Similarly, for the source star, we draw a set of astrometric parameters from a multi-variate Gaussian distribution defined by the source astrometric solution parameter values and their covariance matrix provided in \gaia.}

\item{We use the lens mass estimate $M_{\mbox{\scriptsize L}}$ from Section~\ref{sec:mass}, and the lens and source parallaxes $\varpi_{\mbox{\scriptsize L}}$ and $\varpi_{\mbox{\scriptsize S}}$, respectively, drawn in step~(i), to calculate the Einstein radius $\theta_{\mbox{\scriptsize E}}$:
\begin{equation}
\theta_{\mbox{\scriptsize E}} = \sqrt{ \frac{4\,G M_{\mbox{\scriptsize L}}}{c^{2}} \left( \varpi_{\mbox{\scriptsize L}} - \varpi_{\mbox{\scriptsize S}} \right) }
\label{eqn:theta_e}
\end{equation}
where $G$ is the gravitational constant and $c$ is the speed of light.
\item We compute the path of the source relative to the lens for the time period from the start of the \textit{Gaia} science observations on 25th July $2014$ to 1st January $2070$.}

\item{We compute the lens-to-source flux ratio $f_{\mbox{\scriptsize L}} / f_{\mbox{\scriptsize S}}$ using the faintest lens magnitude from the Pan-STARRS1 $grizy$ passbands, and the source $G$-band magnitude. This minimizes the dilution due to the lens flux in the unresolved microlensing regime, thus maximizing the expected microlensing signals when performing follow-up observations with only the $grizy$ filters available (see Eq. 8 and 12 in Bramich 2018). We also adopt \textit{Gaia}'s resolution of $103$~mas (Fabricius et al. 2016). We use these values along with the relative path computed in step~(iii) to calculate the amplitudes of the photometric and astrometric lensing signals in the unresolved (if necessary) and partially-resolved microlensing regimes (see Section~2 in Bramich 2018).}
\end{enumerate}

We then calculate the median amplitude of each of the microlensing signals over all of the simulations for the source-lens pair. Collecting these results for the 23,253 source-lens pairs, we reject all source-lens pairs for which none of the median microlensing signal amplitudes exceed 0.4~mmag for photometric signals or 0.131~mas for astrometric signals (the same criteria as used in Bramich (2018) and Bramich and Nielsen (2018)). We also reject all source-lens pairs for which the epoch $t_{0}$ of the microlensing event peak falls outside of the time period 1st January 2018 to 1st January 2070.

The final set of predicted microlensing events consists of 27 events caused by 16 unique lens stars.

\begin{table}
\caption{Predicted microlensing events ordered by epoch of the event peak. Microlensing Events (ME) are named in continuation of the events presented in Bramich (2018). Discovery IDs are the identification name given for the initial discovery of the lens object (as given by Best, Magnier, et al. 2018). The details of each event are given with uncertainties in Tables \ref{tab:photevents1}-\ref{tab:astroevents2}. We classify the events into a combination of three categories: photometric (P); high-amplitude astrometric events (HA); and low-amplitude astrometric events (LA).}
\centering
\begin{tabular}{cllrc}
Peak Epoch    & Discovery ID                & Event & Event & Source  \\
(Julian year) & ID                          & Name  & Type  & G-mag.\\
\hline\hline
2019.45       & vB 10 						& ME90  & -/HA  & 20.82\\
2021.31       & 2MASS J05441150-2433018 	& ME94  & -/LA   & 18.67\\ 
2022.87       & vB 10 						& ME101 & -/LA   & 19.80\\
2025.40       & vB 10						& ME100 & -/LA   & 19.77\\
2026.44       & vB 10 						& ME89  & -/HA  & 19.36\\
2027.07       & vB 10           			& ME84  & P/HA  & 20.29\\
2030.37       & GJ 1245B 					& ME103 & -/LA   & 15.61\\
2032.10       & LSR J0011+5908 				& ME93  & -/LA   & 20.13\\
2033.81       & WISE J200050.19+362950.1 	& ME88  & P/HA  & 19.10\\
2036.42       & vB 10 						& ME99  & -/LA   & 20.79\\
2036.93       & VVV BD001            		& ME82  & P/HA  & 20.81\\
2043.84       & WISE J192841.35+235604.9 	& ME87  & P/HA  & 19.01\\
2043.87       & GJ 1245B 					& ME91  & -/HA  & 17.08\\
2044.72       & PSO J076.7092+52.6087		& ME77  & P/HA  & 18.69\\
2045.78       & DENIS-P J0751164-253043 	& ME81  & P/LA  & 20.94\\
2051.71       & 2MASS J15485834-1636018 	& ME96  & -/LA   & 19.67\\
2052.52       & WISEA J053257.29+041842.5 	& ME78  & P/LA  & 17.47\\
2053.76       & VVV BD001 					& ME98  & -/LA   & 17.13\\
2055.35       & vB 10           			& ME86  & P/HA  & 19.64\\
2056.08       & 2MASS J05591914-1404488 	& ME79  & P/HA  & 20.00\\
2057.59       & LHS 3003 					& ME95  & -/LA   & 20.88\\
2061.19       & 2MASS J18284076+1229207 	& ME83  & P/HA  & 17.80\\
2062.26       & WISE J070159.79+632129.2 	& ME80  & P/LA  & 20.57\\
2063.28       & vB 10 						& ME102 & -/LA   & 20.40\\
2064.92       & VVV BD001 					& ME97  & -/LA   & 17.90\\
2065.74       & vB 10          	 			& ME85  & P/HA  & 20.08\\
2068.37       & LSPM J2158+6117 			& ME92  & -/HA  & 17.37\\
\end{tabular}
\label{tab:summary}
\end{table}

\section{Results}
Table \ref{tab:summary} summarizes our predictions of microlensing events by VLM objects up to the year 2070, when comparing \pan and \gaia. In total $27$ events are predicted to occur between 2018 and 2070. We classify the events in terms of either producing a photometric event (P) or not (-), and producing either a high- or low-amplitude astrometric event (HA or LA). The classification is made based on the probability that a source-lens pair will produce either a photometric amplification of the source above $0.4$~mmag for P events, a source deflection above $0.3$~mas for HA events, or a source deflection between $0.131$ and $0.3$~mas for LA events. The probability is derived from the Monte Carlo simulations in Section \ref{sec:monte}, where we require that at least 2.3\% (photometric) and 50\% (astrometric) of the simulations exceed the respective thresholds for a classification to be made. Note that all of our predicted P events are also astrometric events, 9 of them are HA events, while the remaining 3 are just below the limit of $0.3$~mas and are therefore classified as LA events.

Appendix \ref{app:plots} shows how the predicted microlensing events are expected to unfold and the details of each event are listed in Appendix \ref{app:details}. Note that LA events are not included in the appendices, but are available in the online material\footnote{Hosted by the CDS at http://vizier.u-strasbg.fr/viz-bin/VizieR}. 

Out of the 16 unique lenses for which we make predictions, 9 are present in GDR2 (see Appendix B). However, likely due to the fact that they are very faint in the \textit{Gaia} $G$-passband, they fail the selection criteria for source stars adopted by Bramich (2018) and based on the noise and precision criteria presented by Gaia Collaboration, Brown, et al. (2018). They therefore do not appear in the set of predictions by Bramich (2018) and Bramich \& Nielsen (2018).

Below we describe a few noteworthy lens stars that will each produce more than one event in the coming decades.

\subsection{vB 10}
The M8 dwarf star vB 10 (van Biesbroeck 1944) is the most prolific producer of lensing events, with a total of 9 predicted events (ME84-ME86, ME89-ME90, ME99-ME102), the earliest of which (ME90) will peak in 2019, starting mid-2018. Three of these events are P/HA events, and two of these events are -/HA events. The remaining events are -/LA events, and so they will be more difficult to observe. The large number of events is mainly due to the lens star's relative proximity to Earth ($\sim5.95\,$pc), and therefore also its high proper motion ($\sim1488\,\mathrm{mas/yr}$), combined with its location in the crowded Galactic plane. 

Pravdo and Shaklan (2009) suggested that this star might harbor a sub-stellar companion. However, this was later shown not to be the case (Bean et al. 2010; Anglada-Escud\'e et al. 2010; Lazorenko et al. 2011). Observations of the microlensing events present an additional opportunity to detect potential orbiting companion objects.

It should be noted that vB 10 is a component of the multiple-star system WDS J19169+0510$^($\footnote{The primary component, HD 180617, is an M3 dwarf star which is shown by Bramich and Nielsen (2018) to undergo a total of 29 microlensing events up to the end of the 21st century.}$^)$ (Mason et al. 2001). The orbital motions of multiple-star systems are not taken into account in our predictions, and as such they may not be accurate far in the future if the orbital period is comparatively short. Despite the vigorous follow-up observations of vB 10 reported in the literature, we were unable to find estimates of the orbital period for the WDS J19169+0510 system. However, based on the parallax ($0.168\arcsec$) and angular separation ($74\arcsec$) van Biesbroeck (1961) estimated an orbital separation of $\sim440$AU between vB 10 and the primary HD 180617. Assuming masses of $M_{A}\approx0.45M_{\odot}$ (Linsky et al. 1995) and $M_{B}\lesssim0.08$ (Pravdo and Shaklan 2009), this amounts to a lower limit on the orbital period of $\sim12,300$ years. We therefore expect our predictions for vB 10 to be accurate for the next few decades. 
 
\subsection{VVV BD001}
VVV BD001 is a L5 brown dwarf (Beamin et al. 2013), at a distance of $17.54$pc and a proper motion of $640\,\mathrm{mas/yr}$. Despite its likely very low mass and low proper motion compared to vB 10, this star will cause three events (ME82, ME97, ME98). One of these events (ME82) is a potential photometric event in 2036. Apart from yielding a high precision mass measurement for a brown dwarf, this event could probe for planetary companions.

\subsection{GJ 1245B}
GJ 1245B is a flaring M6 dwarf (Giclas, Burnham, and Thomas 1967; Rodono, Ciatti, and Vittone 1980), and so is relatively massive compared to the other stars for which we predict microlensing events. Likewise it is relatively bright with an apparent V-band magnitude of $m_{V}\sim14$. One would therefore expect this to already have been observed by \textit{Gaia}. However, likely due to its high proper motion ($596.4\,\mathrm{mas/yr}$) this star appears to have been missed from \gaia. This star will cause two events (ME91 and ME103), one of which we classify as a -/HA event. ME91 is interesting in that it will produce multiple peaks over a range of years. 

GJ 1245B is the wide tertiary component of a triple system (G208-44 A and B / G208-45). To our knowledge the only estimate of an orbital period comes from Harrington, Dahn, and Guetter (1974), who state that it may be between $400-700$ years. Our predictions suggest GJ 1245B will undergo microlensing events in 2030 and 2043. Since the time until the predicted events are a significant fraction of the estimated orbital period of the star, our predictions of the characteristics of the events may not be accurate. However, we expect that at least the -/HA event in 2043 will still produce a measurable deflection of the source star, as this effect falls off slowly with angular distance from the lens.

\section{Discussion \& Conclusions}
We present predictions of $27$ microlensing events by VLM objects that will occur within the next $50$ years. Of these, $13$ will very likely be high amplitude astrometric events that we expect to be observable by current and future space based observatories. The remaining 14 events will likely require very high precision observations like those from the \textit{Gaia} mission. A total of $12$ are simultaneously photometric and astrometric events. 

The lensing objects found in this work are primarily very late M-dwarfs or brown dwarfs, and so they have either not been observed by \textit{Gaia} or the astrometric parameters are considered unreliable as per Gaia Collaboration, Brown, et al. (2018). However, the Pan-STARRS1 survey provides astrometry of sufficient precision to allow for predictions for the next several decades when compared against the \textit{Gaia} catalog as sources. While future releases of \textit{Gaia} data may improve on these measurements, in which case they should likely be favored over the ground-based Pan-STARRS1 observations, many VLM objects will likely remain unobserved by \textit{Gaia}. This, combined with the continued discoveries of VLM objects by surveys such as Pan-STARRS1 itself, will allow us to make many more predictions of future microlensing events.

\Acknow{MBN is supported by the NYUAD Institute grant G1502. DMB acknowledges the support of the NYU Abu Dhabi Research Enhancement Fund under grant RE124. Many of the calculations performed in this paper employed code from the {\tt DanIDL} library of {\tt IDL} routines (Bramich 2017) available at \url{http://www.danidl.co.uk}. This research was carried out on the High Performance Computing resources at New York University Abu Dhabi. Thanks goes to Nasser Al Ansari, Muataz Al Barwani, Guowei He and Fayizal Mohammed Kunhi for their excellent support. The online material presented in this work is hosted by the Centre de Donn\'es astronomiques de Strasbourg.

This work has made use of data from the European Space Agency (ESA) mission {\it Gaia} (\url{https://www.cosmos.esa.int/gaia}), processed by the {\it Gaia} Data Processing and Analysis Consortium (DPAC, \url{https://www.cosmos.esa.int/web/gaia/dpac/consortium}). Funding for the DPAC has been provided by national institutions, in particular the institutions participating in the {\it Gaia} Multilateral Agreement. 

The Pan-STARRS1 Surveys (PS1) and the PS1 public science archive have been made possible through contributions by the Institute for Astronomy, the University of Hawaii, the Pan-STARRS Project Office, the Max-Planck Society and its participating institutes, the Max Planck Institute for Astronomy, Heidelberg and the Max Planck Institute for Extraterrestrial Physics, Garching, The Johns Hopkins University, Durham University, the University of Edinburgh, the Queen's University Belfast, the Harvard-Smithsonian Center for Astrophysics, the Las Cumbres Observatory Global Telescope Network Incorporated, the National Central University of Taiwan, the Space Telescope Science Institute, the National Aeronautics and Space Administration under Grant No. NNX08AR22G issued through the Planetary Science Division of the NASA Science Mission Directorate, the National Science Foundation Grant No. AST-1238877, the University of Maryland, Eotvos Lorand University (ELTE), the Los Alamos National Laboratory, and the Gordon and Betty Moore Foundation.}

\begin{appendix}
\section{Event paths and expected signals}
\label{app:plots}
\begin{figure}
\centering
\includegraphics[width=\linewidth]{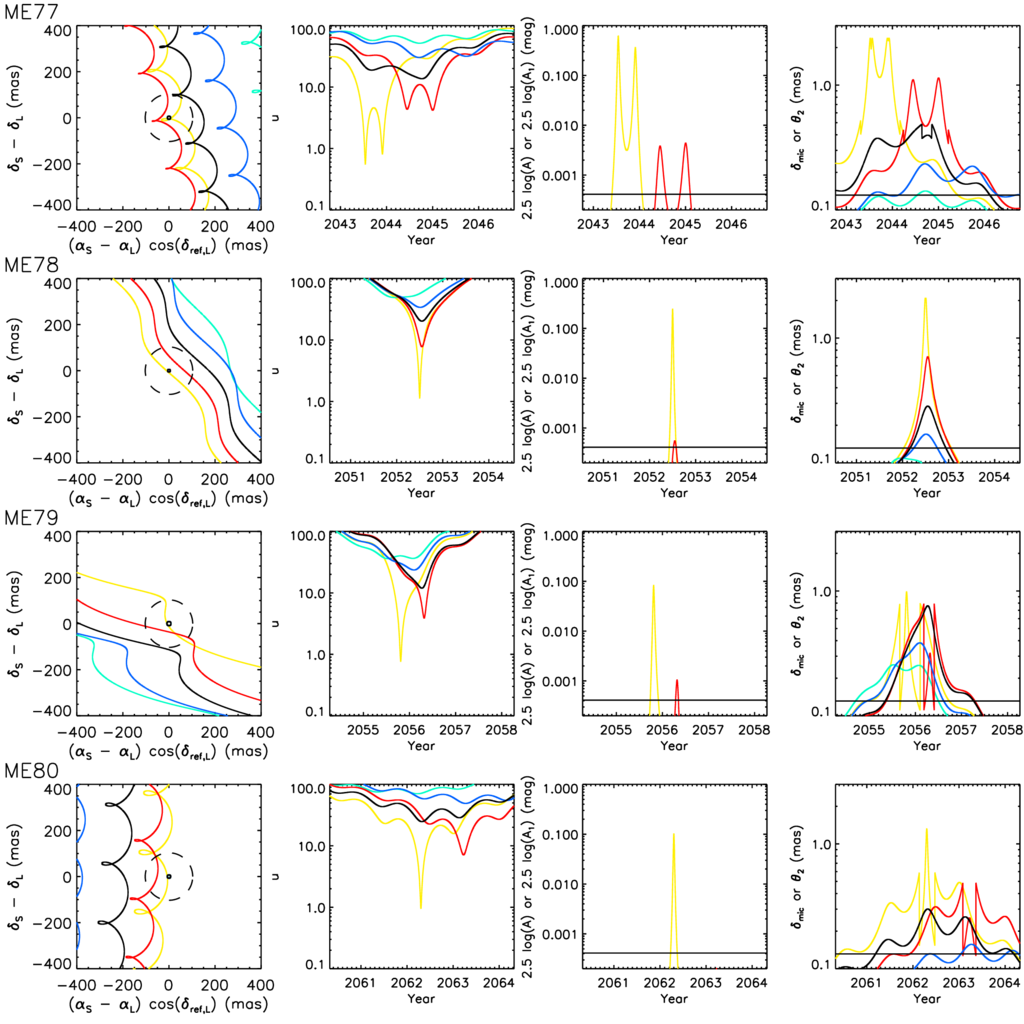}
\caption{Microlensing events ME77-ME80 that exhibit both photometric and astrometric signals. In all panels, five curves are plotted with the colors yellow, red, black, blue, and cyan. Each curve corresponds to the 2.3, 15.9, 50, 84.1, and 97.7 percentiles, respectively, of the results of the Monte Carlo simulations performed in Section~\ref{sec:monte} after they have been ordered by increasing $u_{0}$. \textit{Left-hand panels:} Path of the source star relative to the lens star. The Einstein ring is shown as a solid circle of radius $\theta_{\mbox{\scriptsize E}}$ centered on the lens position. The resolution of \textit{Gaia} (dashed circle) is indicated as a circle of radius 103~mas centered on the lens position. \textit{Middle left-hand panels:} Time-evolution of the normalized source-lens separation $u$.  \textit{Middle right-hand panels:} Time-evolution of the photometric signals $2.5 \log(A)$ (mag; unresolved regime) and $2.5 \log(A_{1})$ (mag; partially-resolved regime). The horizontal black line indicates the photometric precision limit of 0.4~mmag. \textit{Right-hand panels:} Time-evolution of the astrometric signals $\delta_{\mbox{\scriptsize mic}}$ (unresolved regime) and $\theta_{2}$ (partially-resolved regime). The horizontal black line indicates the astrometric precision limit of 0.131~mas.
\label{fig:photevents1}}
\end{figure}

\begin{figure}
\centering
\includegraphics[width=\linewidth]{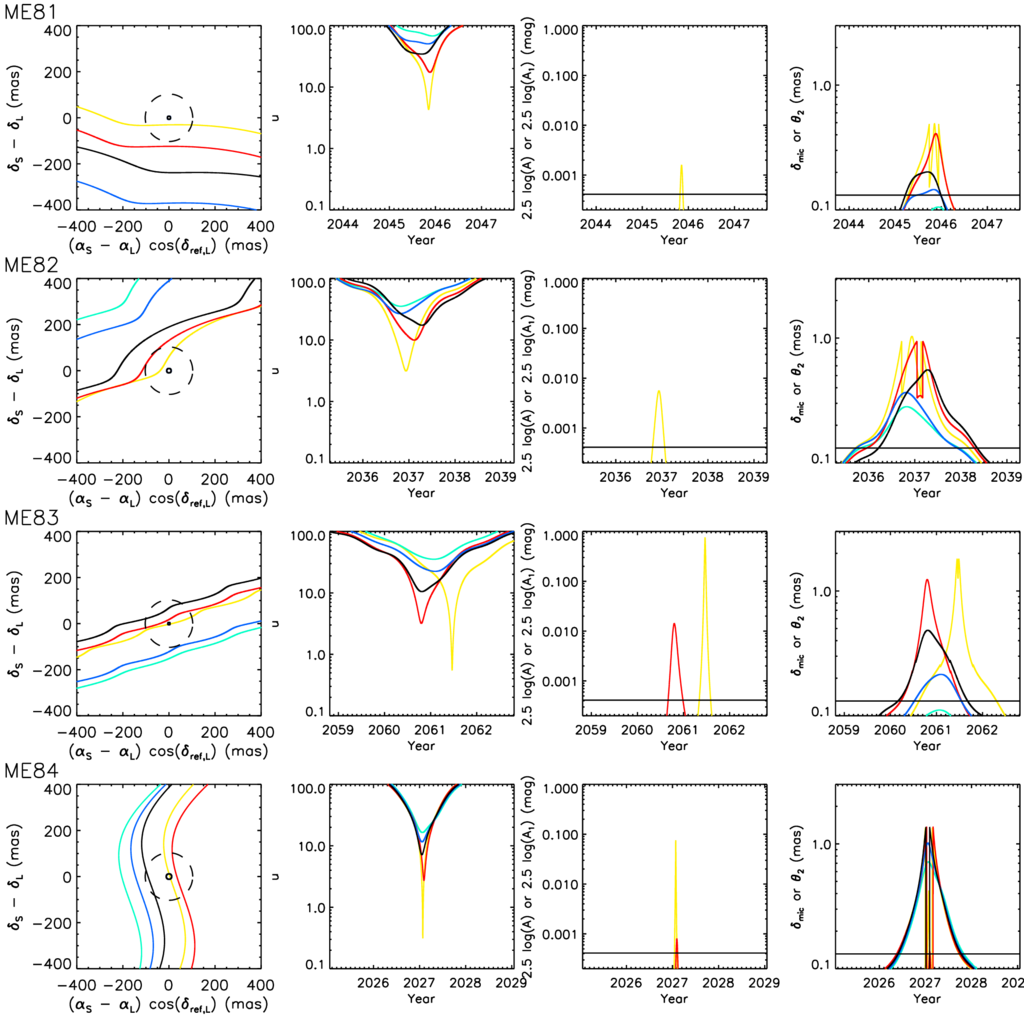}
\caption{Microlensing events ME81-ME84 that exhibit both photometric and astrometric signals. The format of the figure is the same as in Figure~\ref{fig:photevents1}.
 \label{fig:photevents2}}
\end{figure}

\begin{figure}
\centering
\includegraphics[width=\linewidth]{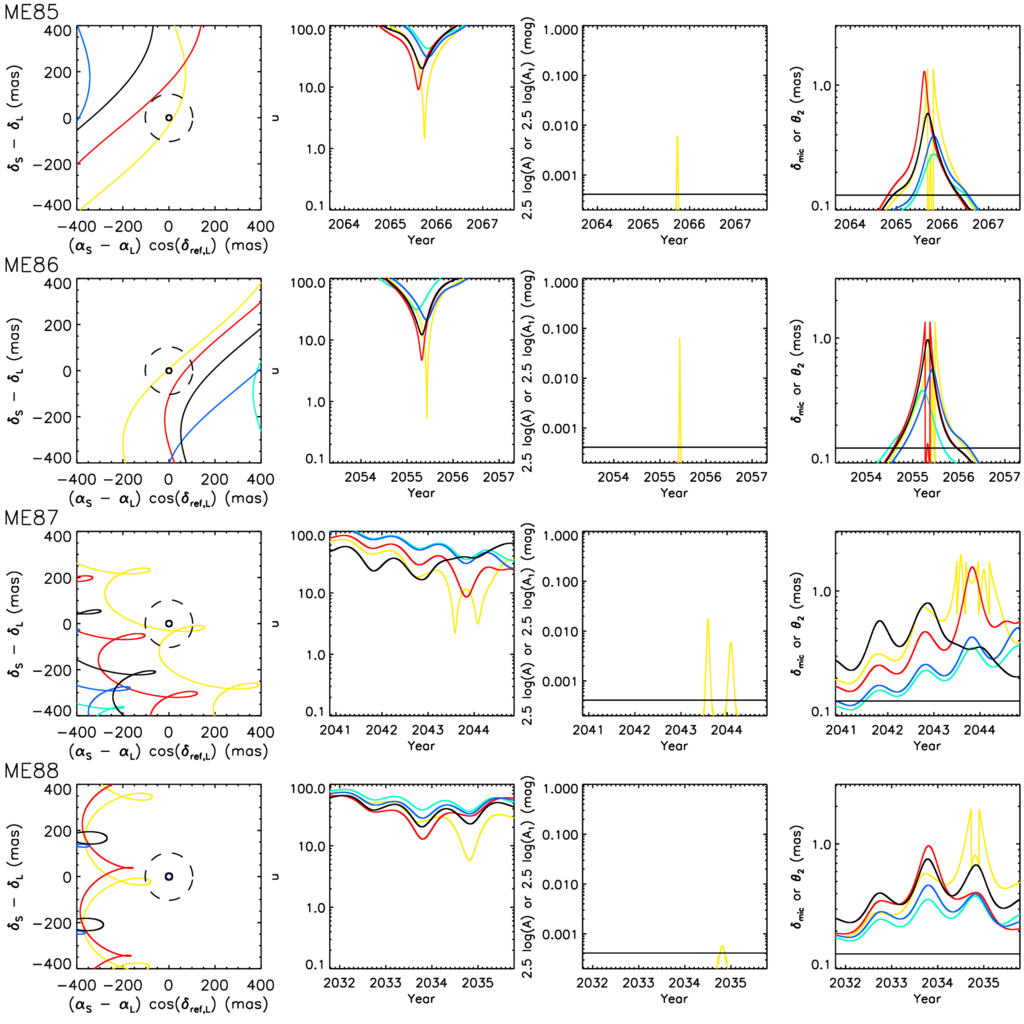}
\caption{Microlensing events ME85-ME88 that exhibit both photometric and astrometric signals. The format of the figure is the same as in Figure~\ref{fig:photevents1}.
\label{fig:photevents3}}
\end{figure}

\begin{figure}
\centering
\includegraphics[width=\linewidth]{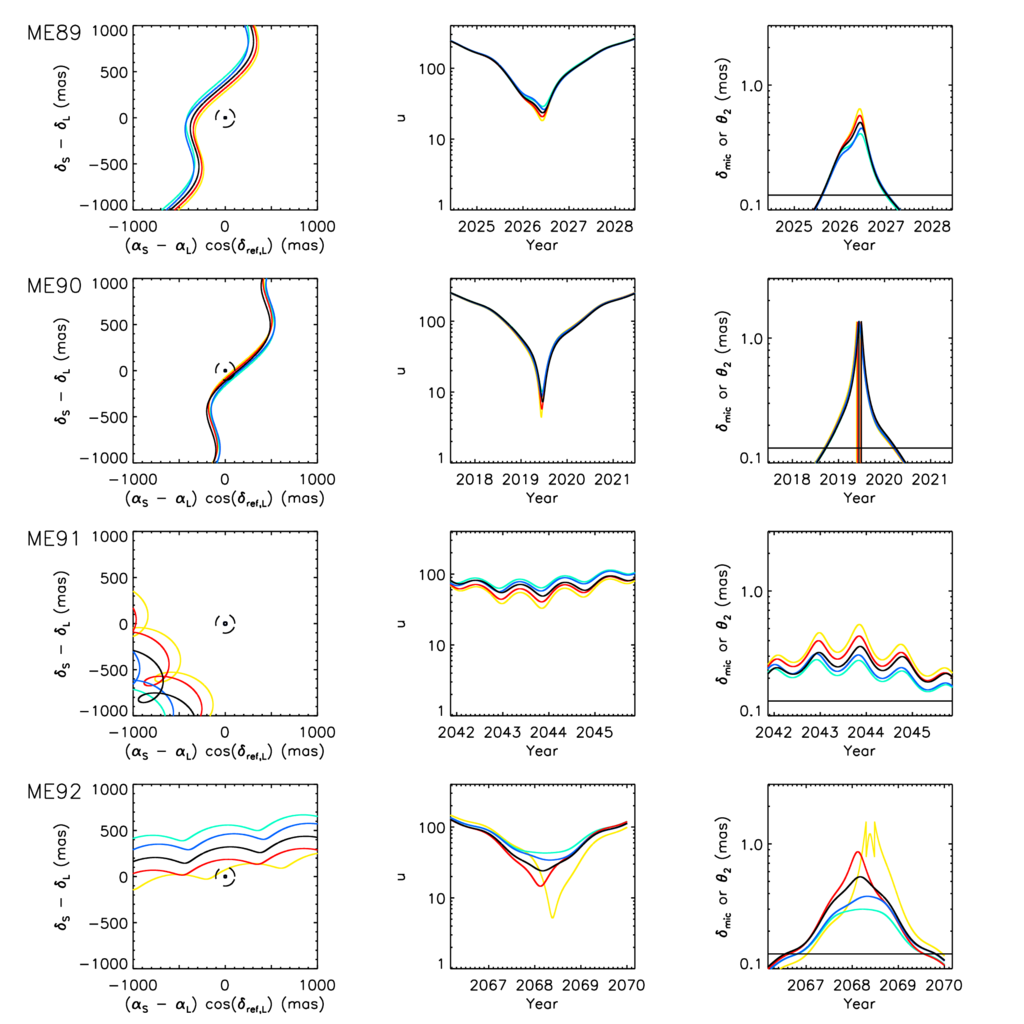}
\caption{Microlensing events ME89-ME92 that exhibit only astrometric signals. The format of the figure is the same as in Figure~\ref{fig:photevents1}, excluding the panels displaying the photometric signals.
\label{fig:astrovents1}}
\end{figure}

\section{Event details}
\label{app:details}
\begin{landscape}
\begin{table}
\centering
\tiny{
\caption{
Characteristics of the photometric microlensing events ME77-ME79 and their constituents. Most quantities are already defined in the text. Spectral types are given as optical/near-IR, as provided by Best, Magnier, et al. (2018). $\Delta(A,A_{1})$ and $\Delta(\delta_{\mbox{\scriptsize mic}},\theta_{2})$ are the differences between the minimum and maximum magnifications and astrometric shifts, respectively, of an event over the time period adopted in this paper. $T[\Delta(A,A_{1})]$ and $T[\Delta(\delta_{\mbox{\scriptsize mic}},\theta_{2})]$ is the amount of time that an event spends with its magnification or astrometric shift above $\min \{ A,A_{1} \} + \Delta(A,A_{1})/2$ and $\min \{ \delta_{\mbox{\scriptsize mic}},\theta_{2} \} + \Delta(\delta_{\mbox{\scriptsize mic}},\theta_{2})/2$, respectively. The numbers in parentheses indicate the uncertainty on the last digit. Passbands marked with * are used in the flux ratio for computing the photometric and astrometric signals (e.g., Eq. 8 and 12 in Bramich 2018). Discovery publications: (a)~Best and Liu (2015), (b)~Kirkpatrick, Schneider, et al. (2014), (c)~Burgasser, Wilson, et al. (2000).
}

\begin{tabular}{@{}l|cc|cc|cc}
\hline
Name                                            & -                     & ME77                         & -                         & ME78                         & -                       & ME79                         \\
\hline
Spectral Type                                   & -/T4.5                & -                            & -/L3                      & -                            & T5/T4.5                 & -                            \\
Discovery ID                                    & PSO J076.7092+52.6087$^{\mathrm{(a)}}$ & -                            & WISEA J053257.29+041842.5$^{\mathrm{(b)}}$ & -                            & 2MASS J05591914-1404488$^{\mathrm{(c)}}$ & -                            \\
PS1 DR1 ID                                      & PSO J076.7092+52.6087 & -                            & PSO J083.2387+04.3117     & -                            & PSO J089.8318-14.0814   & -                            \\
GDR2 Source ID                                  & -                     & 266034294405631232           & 3236465703888892032       & 3236465699593465344          & 2997171394834174976     & 2997170668984296576          \\
$\alpha_{\mbox{\scriptsize ref}}$ (deg$\pm$mas) & 76.709193             & 76.7101065194$\pm$0.323      & 83.238910                 & 83.2420164452$\pm$0.162      & 89.831895               & 89.8390830461$\pm$0.632      \\
$\delta_{\mbox{\scriptsize ref}}$ (deg$\pm$mas) & 52.608725             & 52.6067828658$\pm$0.319      & 4.311572                  & 4.3066772848$\pm$0.164       & $-$14.081459            & $-$14.0856496312$\pm$0.812   \\
Epoch (Julian year)                             & 2010.87847            & 2015.5                       & 2012.36983                & 2015.5                       & 2011.90516              & 2015.5                       \\
$\mu_{\alpha*}$ (mas/year)                      & 57.0(46)              & 2.01(53)                     & 276.2(17)                 & 0.64(28)                     & 569.0(28)               & $-$2.2(14)                   \\
$\mu_{\delta}$ (mas/year)                       & $-$207.2(30)          & $-$0.53(52)                  & $-$440.6(20)              & 0.19(26)                     & $-$339.6(20)            & 1.0(13)                      \\
$\varpi$ (mas)                                  & 59.3(53)              & 0.68(40)                     & 31.6(38)                  & 0.49(19)                     & 96.6(10)                & $-$1.07(82)                  \\
$g_{\mbox{\scriptsize P1}}$ (mag)               & -                     & -                            & -                         & -                            & -                       & -                            \\
$r_{\mbox{\scriptsize P1}}$ (mag)               & -                     & -                            & 21.81(5)*                 & -                            & -                       & -                            \\
$i_{\mbox{\scriptsize P1}}$ (mag)               & -                     & -                            & 19.71(2)                  & -                            & -                       & -                            \\
$z_{\mbox{\scriptsize P1}}$ (mag)               & 20.02(3)*             & -                            & 18.24(1)                  & -                            & 18.02(1)*               & -                            \\
$y_{\mbox{\scriptsize P1}}$ (mag)               & 18.18(3)              & -                            & 17.31(2)                  & -                            & 16.31(2)                & -                            \\
$J$ (2MASS; mag)                                & 15.75(7)              & -                            & 15.44(6)                  & -                            & 13.80(2)                & -                            \\
$H$ (2MASS; mag)                                & 15.35(20)             & -                            & 14.42(6)                  & -                            & 13.68(5)                & -                            \\
$K_{S}$ (2MASS; mag)                            & 15.60(20)             & -                            & 13.86(6)                  & -                            & 13.58(5)                & -                            \\
$W1$ (AllWISE; mag)                             & 14.92(4)              & -                            & 13.30(3)                  & -                            & 13.39(3)                & -                            \\
$W2$ (AllWISE; mag)                             & 13.64(3)              & -                            & 12.97(3)                  & -                            & 11.91(2)                & -                            \\
$W3$ (AllWISE; mag)                             & 11.86(27)             & -                            & 12.13(40)                 & -                            & 10.61(9)                & -                            \\
$W4$ (AllWISE; mag)                             & 8.68                  & -                            & 8.74                      & -                            & 8.74                    & -                            \\
$G$ (mag)                                       & -                     & 18.6940(28)                  & -                         & 17.4666(14)                  & -                       & 20.0007(81)                  \\
$G_{\mbox{\scriptsize BP}}$ (mag)               & -                     & 19.5242(507)                 & -                         & 17.9150(106)                 & -                       & 20.8616(1558)                \\
$G_{\mbox{\scriptsize RP}}$ (mag)               & -                     & 17.8896(294)                 & -                         & 16.8791(73)                  & -                       & 18.9737(494)                 \\
$M$ ($M_{\odot}$)                                & 0.12                  & -                            & 0.12                      & -                            & 0.10                    & -                            \\
$\theta_{\mbox{\scriptsize E}}$ (mas)           & -                     & 7.47(34)                     & -                         & 5.42(35)                     & -                       & 9.08(6)                      \\[3pt]
$u_{0}$ ($\theta_{\mbox{\scriptsize E}}$)       & -                     & 13.67$^{+16.59}_{-9.47}$     & -                         & 20.26$^{+13.70}_{-12.49}$    & -                       & 11.94$^{+11.68}_{-8.05}$     \\[3pt]
$u_{0}$ (mas)                                   & -                     & 101.5$^{+126.6}_{-70.3}$     & -                         & 109.1$^{+72.0}_{-67.1}$      & -                       & 108.4$^{+105.9}_{-73.6}$     \\[3pt]
$t_{0}$ (Julian year)                           & -                     & 2044.72$^{+1.00}_{-0.37}$    & -                         & 2052.516$^{+0.103}_{-0.123}$ & -                       & 2056.084$^{+0.167}_{-0.385}$ \\[3pt]
$\Delta A$ (mag)                                & -                     & 0.0000$^{+0.0043}_{-0.0000}$ & -                         & 0.0000$^{+0.0005}_{-0.0000}$ & -                       & 0.0001$^{+0.0010}_{-0.0000}$ \\[3pt]
$T[\Delta A]$ (d)                               & -                     & 109.7$^{+134.5}_{-68.6}$     & -                         & 82.4$^{+64.6}_{-53.6}$       & -                       & 84.6$^{+76.5}_{-57.0}$       \\[3pt]
$\Delta\theta_{2}$ (mas)                        & -                     & 0.50$^{+0.75}_{-0.27}$       & -                         & 0.27$^{+0.41}_{-0.11}$       & -                       & 0.75$^{+0.04}_{-0.37}$       \\[3pt]
$T[\Delta\theta_{2}]$ (d)                       & -                     & 614$^{+547}_{-456}$          & -                         & 243$^{+206}_{-152}$          & -                       & 227$^{+158}_{-102}$          \\[3pt]
\hline
\end{tabular}
\label{tab:photevents1}
}

\end{table}
\end{landscape}

\begin{landscape}
\begin{table}
\centering
\tiny{
\caption{Characteristics of the photometric microlensing events ME80-ME82 and their constituents. Quantities are the same as in Table~\ref{tab:photevents1}. The numbers in parentheses indicate the uncertainty on the last digit. Discovery publications: (a)~Mace et al. (2013), (b)~Phan-Bao et al. (2008), (c)~Beamin et al. (2013)}
\begin{tabular}{@{}l|cc|cc|cc}
\hline
Name                                            & -                        & ME80                         & -                         & ME81                         & -            & ME82                            \\
\hline
Spectral Type                                   & -/T2.5                   & -                            & L1.5/L1.1                 & -                            & -/L5                     & -                   \\
Discovery ID                                    & WISE J070159.79+632129.2$^{\mathrm{(a)}}$ & -                            & DENIS-P J0751164-253043$^{\mathrm{(b)}}$ & -                            & VVV BD001$^{\mathrm{(c)}}$                & -                   \\
PS1 DR1 ID                                      & PSO J105.4991+63.3579    & -                            & PSO J117.8150-25.5114     & -                            & PSO J261.6672-27.6344    & -                   \\
GDR2 Source ID                                  & -                        & 1099811455149371520          & 5602408602804816768       & 5602408602805833856          & -            & 4059973524898144896             \\
$\alpha_{\mbox{\scriptsize ref}}$ (deg$\pm$mas) & 105.499205               & 105.4985211481$\pm$0.774     & 117.815211                & 117.8059492494$\pm$1.152     & 261.667255   & 261.6629198694$\pm$16.915       \\
$\delta_{\mbox{\scriptsize ref}}$ (deg$\pm$mas) & 63.358112                & 63.3543980192$\pm$0.770      & $-$25.511497              & $-$25.5101733433$\pm$2.067   & $-$27.634409 & $-$27.6366575488$\pm$10.824    \\
Epoch (Julian year)                             & 2010.47422               & 2015.5                       & 2011.44597                & 2015.5                       & 2012.02303   & 2015.5                          \\
$\mu_{\alpha*}$ (mas/year)                      & $-$17.3(31)              & $-$0.6(16)                   & $-$880.7(12)              & $-$3.1(33)                   & $-$548.7(22) & -                               \\
$\mu_{\delta}$ (mas/year)                       & $-$261.7(44)             & $-$3.8(16)                   & 150.1(15)                 & 4.8(32)                      & $-$329.6(44) & -                               \\
$\varpi$ (mas)                                  & 60.5(62)                 & 0.0(12)                      & 56.30(9)                  & $-$2.2(25)                   & 57.0(40)     & -                               \\
$g_{\mbox{\scriptsize P1}}$ (mag)               & -                        & -                            & -                         & -                            & -            & -                               \\
$r_{\mbox{\scriptsize P1}}$ (mag)               & -                        & -                            & 20.06(2)*                 & -                            & 20.14(1)*    & -                               \\
$i_{\mbox{\scriptsize P1}}$ (mag)               & -                        & -                            & 17.57(1)                  & -                            & 18.14(1)     & -                               \\
$z_{\mbox{\scriptsize P1}}$ (mag)               & 19.43(3)*                & -                            & 16.03(1)                  & -                            & 16.36(1)     & -                               \\
$y_{\mbox{\scriptsize P1}}$ (mag)               & 17.98(2)                 & -                            & 15.04(1)                  & -                            & 15.35(1)     & -                               \\
$J$ (2MASS; mag)                                & 15.79(7)                 & -                            & 13.16(2)                  & -                            & 13.40(3)     & -                               \\
$H$ (2MASS; mag)                                & 15.08(11)                & -                            & 12.50(2)                  & -                            & 12.66(3)     & -                               \\
$K_{S}$ (2MASS; mag)                            & 14.88(11)                & -                            & 11.99(2)                  & -                            & 12.23(3)     & -                               \\
$W1$ (AllWISE; mag)                             & 14.19(3)                 & -                            & 11.69(2)                  & -                            & -            & -                               \\
$W2$ (AllWISE; mag)                             & 13.20(3)                 & -                            & 11.44(2)                  & -                            & -            & -                               \\
$W3$ (AllWISE; mag)                             & 11.69                    & -                            & 11.05(13)                 & -                            & -            & -                               \\
$W4$ (AllWISE; mag)                             & 8.82                     & -                            & 9.13                      & -                            & -            & -                               \\
$G$ (mag)                                       & -                        & 20.5713(94)                  & -                         & 20.9383(162)                 & -            & 20.8126(190)                    \\
$G_{\mbox{\scriptsize BP}}$ (mag)               & -                        & 20.8057(820)                 & -                         & 21.5572(3198)                & -            & -                               \\
$G_{\mbox{\scriptsize RP}}$ (mag)               & -                        & 19.7933(483)                 & -                         & 19.6725(782)                 & -            & -                               \\
$M$ ($M_{\odot}$)                                & 0.11                     & -                            & 0.10                      & -                            & 0.21         & -                               \\
$\theta_{\mbox{\scriptsize E}}$ (mas)           & -                        & 7.42(40)                     & -                         & 7.01(15)                     & -            & 9.98(36)                        \\
$u_{0}$ ($\theta_{\mbox{\scriptsize E}}$)       & -                        & 24.75$^{+24.21}_{-17.62}$    & -                         & 34.35$^{+16.08}_{-16.97}$    & -            & 17.38$^{+9.42}_{-7.43}$         \\[3pt]
$u_{0}$ (mas)                                   & -                        & 184$^{+175}_{-131}$          & -                         & 240$^{+110}_{-118}$          & -            & 174.3$^{+91.9}_{-75.8}$         \\[3pt]
$t_{0}$ (Julian year)                           & -                        & 2062.26(99)                  & -                         & 2045.780$^{+0.110}_{-0.313}$ & -            & 2036.9309$^{+0.2585}_{-0.0909}$ \\[3pt]
$\Delta A$ (mag)                                & -                        & 0.0000$^{+0.0002}_{-0.0000}$ & -                         & 0.0000$^{+0.0000}_{-0.0000}$ & -            & 0.0000$^{+0.0001}_{-0.0000}$    \\[3pt]
$T[\Delta A]$ (d)                               & -                        & 127.4$^{+176.0}_{-80.6}$     & -                         & 151.9$^{+85.8}_{-78.3}$      & -            & 146.5$^{+50.2}_{-53.1}$         \\[3pt]
$\Delta\theta_{2}$ (mas)                        & -                        & 0.30$^{+0.26}_{-0.15}$       & -                         & 0.20$^{+0.20}_{-0.07}$       & -            & 0.57$^{+0.33}_{-0.20}$          \\[3pt]
$T[\Delta\theta_{2}]$ (d)                       & -                        & 788$^{+772}_{-480}$          & -                         & 355$^{+119}_{-148}$          & -            & 339$^{+129}_{-132}$             \\[3pt]
\hline
\end{tabular}
\label{tab:photevents2}
}

\end{table}
\end{landscape}

\begin{landscape}
\begin{table}
\centering
\tiny{
\caption{Characteristics of the photometric microlensing events ME83-ME85 and their constituents. Quantities are the same as in Table~\ref{tab:photevents1}. The numbers in parentheses indicate the uncertainty on the last digit. Discovery publications: (a)~Kirkpatrick, Looper, et al. (2010), (b)~van Biesbroech (1961).}
\begin{tabular}{@{}l|cc|cc|cc}
\hline
Name                                            & -                       & ME83                         & -             & ME84                            & -             & ME85                            \\
\hline
Spectral Type                                   & -/M7.5                  & -                            & M8/M8                         & -                               & M8/M8                      & -  \\
Discovery ID                                    & 2MASS J18284076+1229207$^{\mathrm{(a)}}$ & -                            & vB 10$^{\mathrm{(b)}}$                         & -                               & vB 10$^{\mathrm{(b)}}$                      & -  \\
PS1 DR1 ID                                      & PSO J277.1690+12.4888   & -                            & PSO J289.2382+05.1464         & -                               & PSO J289.2382+05.1464      & -  \\
GDR2 Source ID                                  & 4484348145137238016     & 4484348145137243904          & 4293315765165489536           & 4293315765149752320             & 4293315765165489536             & 4293315662070462080             \\
$\alpha_{\mbox{\scriptsize ref}}$ (deg$\pm$mas) & 277.169077              & 277.1656374919$\pm$0.120     & 289.238239    & 289.2355244202$\pm$1.572        & 289.238239    & 289.2291465168$\pm$0.826        \\
$\delta_{\mbox{\scriptsize ref}}$ (deg$\pm$mas) & 12.488828               & 12.4876488409$\pm$0.145      & 5.146342      & 5.1401535039$\pm$1.250          & 5.146342      & 5.1256517599$\pm$0.807          \\
Epoch (Julian year)                             & 2011.31003              & 2015.5                       & 2010.72041    & 2015.5                          & 2010.72041    & 2015.5                          \\
$\mu_{\alpha*}$ (mas/year)                      & $-$244.4(23)            & $-$1.80(28)                  & $-$594.9(22)  & $-$2.3(33)                      & $-$594.9(22)  & $-$9.5(18)                      \\
$\mu_{\delta}$ (mas/year)                       & $-$90.2(14)             & $-$5.59(30)                  & $-$1364.1(19) & $-$5.9(26)                      & $-$1364.1(19) & $-$9.0(18)                      \\
$\varpi$ (mas)                                  & 16.4(27)                & 0.41(16)                     & 168.15(50)    & 0.02(137)                       & 168.15(50)    & 0.27(91)                        \\
$g_{\mbox{\scriptsize P1}}$ (mag)               & -                       & -                            & -             & -                               & -             & -                               \\
$r_{\mbox{\scriptsize P1}}$ (mag)               & 21.40(11)*              & -                            & 16.59(1)*     & -                               & 16.59(1)*     & -                               \\
$i_{\mbox{\scriptsize P1}}$ (mag)               & 18.39(1)                & -                            & -             & -                               & -             & -                               \\
$z_{\mbox{\scriptsize P1}}$ (mag)               & 16.95(1)                & -                            & -             & -                               & -             & -                               \\
$y_{\mbox{\scriptsize P1}}$ (mag)               & 16.22(1)                & -                            & -             & -                               & -             & -                               \\
$J$ (2MASS; mag)                                & 14.61(4)                & -                            & 9.91(3)       & -                               & 9.91(3)       & -                               \\
$H$ (2MASS; mag)                                & 14.05(5)                & -                            & 9.23(3)       & -                               & 9.23(3)       & -                               \\
$K_{S}$ (2MASS; mag)                            & 13.79(6)                & -                            & 8.77(2)       & -                               & 8.77(2)       & -                               \\
$W1$ (AllWISE; mag)                             & 13.46(3)                & -                            & 8.54(2)       & -                               & 8.54(2)       & -                               \\
$W2$ (AllWISE; mag)                             & 13.17(3)                & -                            & 8.32(2)       & -                               & 8.32(2)       & -                               \\
$W3$ (AllWISE; mag)                             & 12.68(51)               & -                            & 8.04(2)       & -                               & 8.04(2)       & -                               \\
$W4$ (AllWISE; mag)                             & 9.09                    & -                            & 8.40(47)      & -                               & 8.40(47)      & -                               \\
$G$ (mag)                                       & -                       & 17.7971(13)                  & -             & 20.2936(105)                    & -             & 20.0838(74)                     \\
$G_{\mbox{\scriptsize BP}}$ (mag)               & -                       & 18.3452(430)                 & -             & -                               & -             & 20.1466(1507)                   \\
$G_{\mbox{\scriptsize RP}}$ (mag)               & -                       & 16.7806(388)                 & -             & -                               & -             & 18.9724(454)                    \\
$M$ ($M_{\odot}$)                                & 0.19                    & -                            & 0.10          & -                               & 0.10          & -                               \\
$\theta_{\mbox{\scriptsize E}}$ (mas)           & -                       & 4.91(42)                     & -             & 11.90(5)                        & -             & 11.89(4)                        \\
$u_{0}$ ($\theta_{\mbox{\scriptsize E}}$)       & -                       & 10.50$^{+11.90}_{-7.30}$     & -             & 7.19$^{+4.42}_{-4.45}$          & -             & 20.00$^{+10.58}_{-10.88}$       \\[3pt]
$u_{0}$ (mas)                                   & -                       & 51.9$^{+58.5}_{-36.2}$       & -             & 85.6$^{+52.3}_{-52.8}$          & -             & 238$^{+127}_{-129}$             \\[3pt]
$t_{0}$ (Julian year)                           & -                       & 2061.190$^{+0.369}_{-0.521}$ & -             & 2027.0700$^{+0.0331}_{-0.0271}$ & -             & 2065.7396$^{+0.0591}_{-0.0654}$ \\[3pt]
$\Delta A$ (mag)                                & -                       & 0.0002$^{+0.0140}_{-0.0002}$ & -             & 0.0002$^{+0.0006}_{-0.0001}$    & -             & 0.0000$^{+0.0001}_{-0.0000}$    \\[3pt]
$T[\Delta A]$ (d)                               & -                       & 85.2$^{+108.7}_{-55.0}$      & -             & 20.7$^{+32.2}_{-8.5}$           & -             & 55.5$^{+24.1}_{-28.6}$          \\[3pt]
$\Delta\theta_{2}$ (mas)                        & -                       & 0.44$^{+0.82}_{-0.23}$       & -             & 1.35$^{+0.02}_{-0.33}$          & -             & 0.59$^{+0.70}_{-0.21}$          \\[3pt]
$T[\Delta\theta_{2}]$ (d)                       & -                       & 269$^{+267}_{-174}$          & -             & 75.3$^{+71.8}_{-16.6}$          & -             & 151.8$^{+75.6}_{-78.6}$         \\[3pt]
\hline
\end{tabular}
\label{tab:photevents3}
}

\end{table}
\end{landscape}

\begin{landscape}
\begin{table}
\centering
\tiny{
\caption{Characteristics of the photometric microlensing events ME86-ME88 and their constituents. Quantities are the same as in Table~\ref{tab:photevents1}. The numbers in parentheses indicate the uncertainty on the last digit. Discovery publications: (a)~van Biesbroeck (1961), (b)~Mace et al. (2013), (c)~Cushing et al. (2014).}
\begin{tabular}{@{}l|cc|cc|cc}
\hline
Name                                            & -             & ME86                         & -                        & ME87                            & -                        & ME88                            \\
\hline
Spectral Type                                   & M8/M8                 & -                    & T6                       & -                               & T8                       & -                               \\
Discovery ID                                    & vB 10$^{\mathrm{(a)}}$                 & -                    & WISE J192841.35+235604.9$^{\mathrm{(b)}}$ & -                               & WISE J200050.19+362950.1$^{\mathrm{(c)}}$ & -                               \\
PS1 DR1 ID                                      & PSO J289.2382+05.1464 & -                    & PSO J292.1721+23.9346    & -                               & PSO J300.2091+36.4974    & -                               \\
GDR2 Source ID                                  & 4293315765165489536   & 4293315662070489088          & -                        & 2019802550989835648             & -                        & 2060116591502525952             \\
$\alpha_{\mbox{\scriptsize ref}}$ (deg$\pm$mas) & 289.238239    & 289.2309372394$\pm$0.569     & 292.172026               & 292.1697862526$\pm$0.170        & 300.209082               & 300.2090660979$\pm$0.216        \\
$\delta_{\mbox{\scriptsize ref}}$ (deg$\pm$mas) & 5.146342      & 5.1294532938$\pm$0.537       & 23.934761                & 23.9369125251$\pm$0.265         & 36.497219                & 36.4997585371$\pm$0.257         \\
Epoch (Julian year)                             & 2010.72041    & 2015.5                       & 2012.57697               & 2015.5                          & 2009.84657               & 2015.5                          \\
$\mu_{\alpha*}$ (mas/year)                      & $-$594.9(22)  & $-$2.7(13)                   & $-$233.5(41)             & $-$4.10(50)                     & 15.8(45)                 & 1.74(51)                        \\
$\mu_{\delta}$ (mas/year)                       & $-$1364.1(19) & $-$4.1(12)                   & 246.6(36)                & $-$10.15(58)                    & 370.5(51)                & $-$4.33(58)                     \\
$\varpi$ (mas)                                  & 168.15(50)    & 0.41(57)                     & 146.6(129)               & $-$0.16(40)                     & 140.5(124)               & 0.36(30)                        \\
$g_{\mbox{\scriptsize P1}}$ (mag)               & -             & -                            & -                        & -                               & -                        & -                               \\
$r_{\mbox{\scriptsize P1}}$ (mag)               & 16.59(1)*     & -                            & -                        & -                               & -                        & -                               \\
$i_{\mbox{\scriptsize P1}}$ (mag)               & -             & -                            & -                        & -                               & -                        & -                               \\
$z_{\mbox{\scriptsize P1}}$ (mag)               & -             & -                            & 18.43(6)*                & -                               & -                        & -                               \\
$y_{\mbox{\scriptsize P1}}$ (mag)               & -             & -                            & 16.84(1)                 & -                               & 18.38(2)*                & -                               \\
$J$ (2MASS; mag)                                & 9.91(3)       & -                            & 14.34(6)                 & -                               & 15.81(9)                 & -                               \\
$H$ (2MASS; mag)                                & 9.23(3)       & -                            & 14.31(6)                 & -                               & 15.81                    & -                               \\
$K_{S}$ (2MASS; mag)                            & 8.77(2)       & -                            & 14.09(6)                 & -                               & 16.80                    & -                               \\
$W1$ (AllWISE; mag)                             & 8.54(2)       & -                            & 13.57(3)                 & -                               & 15.08(6)                 & -                               \\
$W2$ (AllWISE; mag)                             & 8.32(2)       & -                            & 12.02(2)                 & -                               & 12.69(3)                 & -                               \\
$W3$ (AllWISE; mag)                             & 8.04(2)       & -                            & 10.80(9)                 & -                               & 11.21(10)                & -                               \\
$W4$ (AllWISE; mag)                             & 8.40(47)      & -                            & 8.65                     & -                               & 9.42                     & -                               \\
$G$ (mag)                                       & -             & 19.6399(54)                  & -                        & 19.0056(22)                     & -                        & 19.0999(25)                     \\
$G_{\mbox{\scriptsize BP}}$ (mag)               & -             & 20.2794(756)                 & -                        & 19.8737(409)                    & -                        & 19.5940(373)                    \\
$G_{\mbox{\scriptsize RP}}$ (mag)               & -             & 18.5282(436)                 & -                        & 17.7383(114)                    & -                        & 18.1739(253)                    \\
$M$ ($M_{\odot}$)                                & 0.10          & -                            & 0.15                     & -                               & 0.17                     & -                               \\
$\theta_{\mbox{\scriptsize E}}$ (mas)           & -             & 11.89(3)                     & -                        & 13.21(58)                       & -                        & 13.90(64)                       \\[3pt]
$u_{0}$ ($\theta_{\mbox{\scriptsize E}}$)       & -             & 12.14$^{+9.03}_{-7.43}$      & -                        & 16.37$^{+8.84}_{-7.80}$         & -                        & 20.35$^{+7.81}_{-7.50}$         \\[3pt]
$u_{0}$ (mas)                                   & -             & 144$^{+108}_{-89}$           & -                        & 214$^{+117}_{-98}$              & -                        & 285(102)                        \\[3pt]
$t_{0}$ (Julian year)                           & -             & 2055.3468(509)               & -                        & 2043.8385$^{+0.0359}_{-1.0095}$ & -                        & 2033.8048$^{+1.0119}_{-0.0151}$ \\[3pt]
$\Delta A$ (mag)                                & -             & 0.0000$^{+0.0002}_{-0.0000}$ & -                        & 0.0000$^{+0.0002}_{-0.0000}$    & -                        & 0.0000$^{+0.0000}_{-0.0000}$    \\[3pt]
$T[\Delta A]$ (d)                               & -             & 37.8$^{+23.1}_{-25.2}$       & -                        & 115.0$^{+111.6}_{-38.6}$        & -                        & 113.8$^{+98.7}_{-24.8}$         \\[3pt]
$\Delta\theta_{2}$ (mas)                        & -             & 0.97$^{+0.38}_{-0.41}$       & -                        & 0.79$^{+0.68}_{-0.28}$          & -                        & 0.67$^{+0.42}_{-0.20}$          \\[3pt]
$T[\Delta\theta_{2}]$ (d)                       & -             & 103.4$^{+63.7}_{-52.2}$      & -                        & 574$^{+428}_{-362}$             & -                        & 656$^{+382}_{-315}$             \\[3pt]
\hline
\end{tabular}
\label{tab:photevents4}
}

\end{table}
\end{landscape}

\begin{landscape}
\begin{table}
\centering
\tiny{
\caption{Characteristics of the astrometric microlensing events ME89-ME91 and their constituents. Quantities are the same as in Table~\ref{tab:photevents1}. The numbers in parentheses indicate the uncertainty on the last digit. Discovery publications: (a)~van Biesbroeck (1961), (b)~Giclas, Burnham, and Thomas (1967) and Rodono, Ciatti, and Vittone (1980).}
\begin{tabular}{@{}l|cc|cc|cc}
\hline
Name                                            & -             & ME89                            & -             & ME90                               & -            & ME91                            \\
\hline
Spectral Type                                   & M8/M8         & -                               & M8/M8         & -                  & M6/M6             & -                  \\
Discovery ID                                    & vB 10 $^{\mathrm{(a)}}$             & -                       & vB 10 $^{\mathrm{(a)}}$                     & -                  & GJ 1245B $^{\mathrm{(b)}}$           & -                  \\
PS1 DR1 ID                                      & PSO J289.2382+05.1464 & -                       & PSO J289.2382+05.1464         & -                  & PSO J298.4811+44.4136     & -                  \\
GDR2 Source ID                                  & 4293315765165489536   & 4293315760821832064     & 4293315765165489536           & 4293315765152355456                & -            & 2079073928612822016             \\
$\alpha_{\mbox{\scriptsize ref}}$ (deg$\pm$mas) & 289.238239    & 289.2356137076$\pm$0.563        & 289.238239    & 289.2368286564$\pm$5.342           & 298.480993   & 298.4854577869$\pm$0.082        \\
$\delta_{\mbox{\scriptsize ref}}$ (deg$\pm$mas) & 5.146342      & 5.1404558630$\pm$0.547          & 5.146342      & 5.1430208619$\pm$4.314             & 44.413824    & 44.4090827669$\pm$0.091         \\
Epoch (Julian year)                             & 2010.72041    & 2015.5                          & 2010.72041    & 2015.5                             & 2009.16791   & 2015.5                          \\
$\mu_{\alpha*}$ (mas/year)                      & $-$594.9(22)  & $-$3.9(15)                      & $-$594.9(22)  & -                                  & 357.4(44)    & $-$2.07(19)                     \\
$\mu_{\delta}$ (mas/year)                       & $-$1364.1(19) & $-$3.3(14)                      & $-$1364.1(19) & -                                  & $-$477.5(26) & 2.39(19)                        \\
$\varpi$ (mas)                                  & 168.15(50)    & 2.24(59)                        & 168.15(50)    & -                                  & 220.2(15)    & 1.12(10)                        \\
$g_{\mbox{\scriptsize P1}}$ (mag)               & -             & -                               & -             & -                                  & 14.75(1)*    & -                               \\
$r_{\mbox{\scriptsize P1}}$ (mag)               & 16.59(1)*     & -                               & 16.59(1)*     & -                                  & -            & -                               \\
$i_{\mbox{\scriptsize P1}}$ (mag)               & -             & -                               & -             & -                                  & -            & -                               \\
$z_{\mbox{\scriptsize P1}}$ (mag)               & -             & -                               & -             & -                                  & -            & -                               \\
$y_{\mbox{\scriptsize P1}}$ (mag)               & -             & -                               & -             & -                                  & -            & -                               \\
$J$ (2MASS; mag)                                & 9.91(3)       & -                               & 9.91(3)       & -                                  & 8.28(3)      & -                               \\
$H$ (2MASS; mag)                                & 9.23(3)       & -                               & 9.23(3)       & -                                  & 7.73(3)      & -                               \\
$K_{S}$ (2MASS; mag)                            & 8.77(2)       & -                               & 8.77(2)       & -                                  & 7.39(2)      & -                               \\
$W1$ (AllWISE; mag)                             & 8.54(2)       & -                               & 8.54(2)       & -                                  & 7.12(14)     & -                               \\
$W2$ (AllWISE; mag)                             & 8.32(2)       & -                               & 8.32(2)       & -                                  & 6.93(6)      & -                               \\
$W3$ (AllWISE; mag)                             & 8.04(2)       & -                               & 8.04(2)       & -                                  & 6.79(2)      & -                               \\
$W4$ (AllWISE; mag)                             & 8.40(47)      & -                               & 8.40(47)      & -                                  & 6.66(9)      & -                               \\
$G$ (mag)                                       & -             & 19.3637(54)                     & -             & 20.8212(152)                       & -            & 17.0778(9)                      \\
$G_{\mbox{\scriptsize BP}}$ (mag)               & -             & 19.7712(738)                    & -             & 21.3876(1421)                      & -            & 17.7940(101)                    \\
$G_{\mbox{\scriptsize RP}}$ (mag)               & -             & 18.0230(363)                    & -             & 19.6953(1093)                      & -            & 16.2248(58)                     \\
$M$ ($M_{\odot}$)                                & 0.10          & -                               & 0.10          & -                                  & 0.18         & -                               \\
$\theta_{\mbox{\scriptsize E}}$ (mas)           & -             & 11.82(3)                        & -             & 11.90(2)                           & -            & 17.69(6)                        \\[3pt]
$u_{0}$ ($\theta_{\mbox{\scriptsize E}}$)       & -             & 23.51$^{+2.63}_{-2.85}$         & -             & 7.27$^{+1.39}_{-1.51}$             & -            & 48.99$^{+7.52}_{-8.46}$         \\[3pt]
$u_{0}$ (mas)                                   & -             & 278.1$^{+30.7}_{-33.9}$         & -             & 86.5$^{+16.7}_{-18.0}$             & -            & 867$^{+133}_{-150}$             \\[3pt]
$t_{0}$ (Julian year)                           & -             & 2026.4419$^{+0.0216}_{-0.0230}$ & -             & 2019.44668$^{+0.00903}_{-0.00876}$ & -            & 2043.8475$^{+0.0192}_{-0.9136}$ \\[3pt]
$\Delta\theta_{2}$ (mas)                        & -             & 0.500$^{+0.069}_{-0.051}$       & -             & 1.353$^{+0.005}_{-0.009}$          & -            & 0.343$^{+0.075}_{-0.048}$       \\[3pt]
$T[\Delta\theta_{2}]$ (d)                       & -             & 269.3$^{+32.4}_{-42.9}$         & -             & 49.2$^{+16.0}_{-4.8}$              & -            & 1524$^{+321}_{-293}$            \\ [3pt]
\hline
\end{tabular}
\label{tab:astroevents1}
}
\end{table}
\end{landscape}

\begin{landscape}
\begin{table}
\centering
\tiny{
\caption{Characteristics of the astrometric microlensing event ME92 and its constituents. Quantities are the same as in Table~\ref{tab:photevents1}. The numbers in parentheses indicate the uncertainty on the last digit. Discovery publications: (a)~L\'epine, Rich, and Shara (2003).}
\begin{tabular}{@{}l|cc}
\hline
Name                                            & -                & ME92                         \\
\hline
Spectral Type                                   & M6/-                   & -                      \\
Discovery ID                                    & LSPM J2158+6117 $^{\mathrm{(a)}}$       & -                      \\
PS1 DR1 ID                                      & PSO J329.6504+61.2854  & -                      \\
GDR2 Source ID                                  & 2204001126955518592    & 2204001053930423296          \\
$\alpha_{\mbox{\scriptsize ref}}$ (deg$\pm$mas) & 329.649600       & 329.6763314791$\pm$0.083     \\
$\delta_{\mbox{\scriptsize ref}}$ (deg$\pm$mas) & 61.285353        & 61.2872378952$\pm$0.084      \\
Epoch (Julian year)                             & 2011.48110       & 2015.5                       \\
$\mu_{\alpha*}$ (mas/year)                      & 811.4(21)        & $-$3.03(18)                  \\
$\mu_{\delta}$ (mas/year)                       & 112.2(21)        & $-$2.42(18)                  \\
$\varpi$ (mas)                                  & 59.2(22)         & 0.225(93)                    \\
$g_{\mbox{\scriptsize P1}}$ (mag)               & 17.45(2)*        & -                            \\
$r_{\mbox{\scriptsize P1}}$ (mag)               & 16.17(1)         & -                            \\
$i_{\mbox{\scriptsize P1}}$ (mag)               & -                & -                            \\
$z_{\mbox{\scriptsize P1}}$ (mag)               & -                & -                            \\
$y_{\mbox{\scriptsize P1}}$ (mag)               & 12.63(1)         & -                            \\
$J$ (2MASS; mag)                                & 11.29(3)         & -                            \\
$H$ (2MASS; mag)                                & 10.79(3)         & -                            \\
$K_{S}$ (2MASS; mag)                            & 10.45(2)         & -                            \\
$W1$ (AllWISE; mag)                             & 10.25(2)         & -                            \\
$W2$ (AllWISE; mag)                             & 10.01(2)         & -                            \\
$W3$ (AllWISE; mag)                             & 9.75(5)          & -                            \\
$W4$ (AllWISE; mag)                             & 8.83             & -                            \\
$G$ (mag)                                       & -                & 17.3697(13)                  \\
$G_{\mbox{\scriptsize BP}}$ (mag)               & -                & 18.1874(123)                 \\
$G_{\mbox{\scriptsize RP}}$ (mag)               & -                & 16.4780(58)                  \\
$M$ ($M_{\odot}$)                                & 0.33             & -                            \\
$\theta_{\mbox{\scriptsize E}}$ (mas)           & -                & 12.59(23)                    \\[3pt]
$u_{0}$ ($\theta_{\mbox{\scriptsize E}}$)       & -                & 24.10$^{+10.05}_{-9.51}$     \\[3pt]
$u_{0}$ (mas)                                   & -                & 305$^{+123}_{-121}$          \\[3pt]
$t_{0}$ (Julian year)                           & -                & 2068.374$^{+0.171}_{-0.211}$ \\[3pt]
$\Delta\theta_{2}$ (mas)                        & -                & 0.517$^{+0.339}_{-0.153}$    \\[3pt]
$T[\Delta\theta_{2}]$ (d)                       & -                & 487$^{+176}_{-202}$          \\[3pt]
\hline
\end{tabular}
\label{tab:astroevents2}
}

\end{table}
\end{landscape}

\end{appendix}


\begin{references}
\refitem{Anglada-Escud\'e G., Shkolnik E.~L., Weinberger A.~J., Thompson I.~B., Osip D.~J. \& Debes J.~H.}{2010}{\apj}{711}{L24-L29}
\refitem{Beamin J.~C., Minniti D., Gromadzki M., Kurtev R., Ivanov V.~D., Beletsky Y., Lucas P., Saito R.~K. \& Borissova J.}{2013}{\aap}{557}{L8}
\refitem{Bean J.~L., Seifahrt A., Hartman H., Nilsson H., Reiners A., Dreizler S., Henry T.~J. \& Wiedemann G.}{2010}{\apjl}{711}{L19-L23}
\refitem{Beaulieu, J.-P. et al}{2006}{\nat}{439}{437-440}
\refitem{Belokurov V.~A. \& Evans N.~W.}{2002}{\mnras}{331}{649-665}
\refitem{Best W.~M.~J. et al.}{2015}{\apj}{814}{118}
\refitem{Best W.~M.~J. et al.}{2018}{\apjs}{234}{1}
\refitem{Bramich D.~M.}{2017}{Astrophysics Source Code Library}{record ascl:1709.005}{}
\refitem{Bramich D.~M.}{2018}{A\&A}{Accepted}{B18}
\refitem{Bramich D.~M. \& Nielsen, M.~B.}{2018}{\actaa}{~}{~}
\refitem{Burgasser A.~J., Burrows A. \& Kirkpatrick J.~D.}{2006}{\apj}{639}{1095-1113}
\refitem{Burgasser A.~J. et al.}{2000}{\aj}{120}{1100-1105}
\refitem{Burrows A., Sudarsky D. \& Hubeny I.}{2006}{\apj}{640}{1063-1077}
\refitem{Chambers K.~C. et al.}{2016}{ArXiv e-prints}{1612.05560}{~}
\refitem{Choi J., Dotter A., Conroy C., Cantiello M., Paxton B. \& Johnson B.~D.}{2016}{\apj}{823}{102}
\refitem{Cushing M.~C., Kirkpatrick J.~D., Gelino C.~R., Mace G.~N., Skrutskie M.~F. \& Gould A.}{2014}{\aj}{147}{113}
\refitem{Dominik M. \& Sahu K.~C.}{2000}{\apj}{534}{213-226}
\refitem{Dotter A.}{2016}{\apjs}{222}{8}
\refitem{Dupuy T.~J. \& Liu M.~C.}{2012}{\apjs}{201}{19}
\refitem{Einstein A.}{1915}{Sitzungsber.~preuss.Akad.~Wiss.}{47}{831-839}
\refitem{Einstein A.}{1936}{Science}{84}{506-507}
\refitem{Fabricius C. et al.}{2016}{\aap}{595}{A3}
\refitem{Foreman-Mackey D., Hogg D.~W., Lang D. \& Goodman J.}{2013}{\pasp}{125}{306-312}
\refitem{Gaia Collaboration et al.}{2018}{\aap}{616}{A10}
\refitem{Gaia Collaboration et al.}{2018}{\aap}{616}{A1}
\refitem{Giclas H.~L., Burnham R. \& Thomas N.~G.}{1967}{Lowell Observatory Bulletin}{7}{~}
\refitem{Harrington R.~S., Dahn C.~C. \& Guetter H.~H.}{1974}{\apjl}{194}{L87}
\refitem{Irwin J., Berta-Thompson Z.~K., Charbonneau D., Dittmann J. \& Newton E.~R.}{2015}{American Astronomical Society Meeting Abstracts \#225}{225}{258.01}
\refitem{Jordi C., Gebran M., Carrasco J.~M., de Bruijne J., Voss H., Fabricius C., Knude J., Vallenari A., Kohley R. \& Mora A.}{2010}{\aap}{523}{A48}
\refitem{Kaiser N., Burgett W., Chambers K., Denneau L., Heasley J., Jedicke R., Magnier E., Morgan J., Onaka P. \& Tonry J.}{2010}{Ground-based and Airborne Telescopes III}{7733}{77330E}
\refitem{Kirkpatrick J.~D. et al.}{2010}{\apjs}{190}{100-146}
\refitem{Kirkpatrick J.~D. et al.}{2014}{\apj}{783}{122}
\refitem{Lazorenko P.~F. et al.}{2011}{\aap}{527}{A25}
\refitem{L\'epine S., Rich R.~M. \& Shara M.~M.}{2003}{\aj}{125}{1598-1622}
\refitem{Liebes S.}{1964}{Physical Review}{133}{835-844}
\refitem{Linsky J.~L., Wood B.~E., Brown A., Giampapa M.~S. \& Ambruster C.}{1995}{\apj}{455}{670}
\refitem{Mace G.~N. et al.}{2013}{\apjs}{205}{6}
\refitem{Mason Brian D., Wycoff Gary L., Hartkopf William I., Douglass Geoffrey G. \& Worley Charles E.}{2001}{\aj}{122}{3466-3471}
\refitem{Morton T.D.}{2015}{Astrophysics Source Code Library}{ascl:1503.010}{~}
\refitem{Phan-Bao N. et al.}{2008}{\mnras}{383}{831-844}
\refitem{Pravdo S.~H. \& Shaklan S.~B.}{2009}{\apj}{700}{623-632}
\refitem{Prusti T. et al.}{2016}{\aap}{595}{A1}
\refitem{Refsdal S.}{1964}{\mnras}{128}{295}
\refitem{Rodono M., Ciatti F. \& Vittone A.}{1980}{\aj}{85}{298-301}
\refitem{Rybicki K.~A., Wyrzykowski \L., Klencki J., de Bruijne J., Belczy{\'n}ski K. \& Chru{\'s}li{\'n}ska M.}{2018}{\mnras}{476}{2013-2028}
\refitem{Sahu K.~C. et al.}{2017}{Science}{356}{1046-1050}
\refitem{Sumi T. et al.}{2013}{\apj}{778}{150}
\refitem{Udalski A.}{2003}{\actaa}{53}{291-305}
\refitem{Udalski A., Szyma{\'n}ski M.~K. \& Szyma{\'n}ski G.}{2015}{\actaa}{65}{1-38}
\refitem{van Biesbroeck G.}{1944}{\aj}{{51}}{61}
\refitem{van Biesbroeck G.}{1961}{\aj}{{66}}{528-530}
\refitem{Wyrzykowski \L. et al.}{2015}{\apjs}{216}{12}
\end{references}
\end{document}